\documentclass[prl,twocolumn,superscriptaddress,showpacs]{revtex4-1}

\usepackage{graphicx}
\usepackage{dcolumn}
\usepackage{bm}
\usepackage{accents}

\newcommand{\beq}{\begin{equation}}
\newcommand{\eeq}{\end{equation}}
\newcommand{\beqa}{\begin{eqnarray}}
\newcommand{\eeqa}{\end{eqnarray}}
\newcommand{\Tr}{\text{Tr}}
\newcommand{\del}[2]{\frac{\dd #1}{\dd #2}}

\newcommand{\half}{\frac{1}{2}}

\newcommand{\av}[1]{\left\langle #1 \right\rangle}

\newcommand{\ee}{\mathrm{e}}
\newcommand{\dd}{\mathrm{d}}

\begin{document}


\title{Work fluctuation-dissipation trade-off in heat engines}


\author{Ken Funo}
\email[]{funo@cat.phys.s.u-tokyo.ac.jp}
\affiliation{Department of Physics, The University of Tokyo, 7-3-1 Hongo, Bunkyo-ku, Tokyo 113-0033, Japan}
\author{Masahito Ueda}
\affiliation{Department of Physics, The University of Tokyo, 7-3-1 Hongo, Bunkyo-ku, Tokyo 113-0033, Japan}
\affiliation{RIKEN Center for Emergent Matter Science (CEMS), Wako, Saitama 351-0198, Japan}

\date{\today}
\begin{abstract}
Reducing work fluctuation and dissipation in heat engines or, more generally, information heat engines that perform feedback control is vital to maximize their efficiency. The same problem arises when we attempt to maximize the efficiency of a given thermodynamic task that undergoes nonequilibrium processes for arbitrary initial and final states. We find that the most general trade-off relation between work fluctuation and dissipation applicable to arbitrary nonequilibrium processes is bounded from below by the information distance characterizing how far the system is from thermal equilibrium. The minimum amount of dissipation is found to be given in terms of the relative entropy and the Renyi divergence, both of which quantify the information distance between the state of the system and the canonical distribution. We give an explicit protocol that achieves the fundamental lower bound of the trade-off relation.
\end{abstract}


\maketitle

Recent developments in nonequilibrium statistical mechanics enable us to assign physical meanings to nonequilibrium entropies such as Shannon and von Neumann entropies in certain situations~\cite{Esposito2,Deffner,Parrondo}. The information-theoretic analysis of thermodynamics starting from and ending at arbitrary nonequilibrium states has been carried out, as in encoding and erasure of information~\cite{Landauer,Hasegawa,Esposito2,Berut}. An important subset of this category is the information heat engines~\cite{Parrondo,JMaxwell,Maxwell,Toyabe,Koski1,Koski2,Sagawa1,Maruyama,Sagawa2,Sagawa3}, since the measurement projects the state of the system into the postmeasurement state which is usually out of equilibrium. They play a pivotal role in controlling small thermodynamic systems that operate at the level of thermodynamic fluctuations. Viewing biological processes as information processing requires us to quantify thermodynamic costs of biological sensory adaptation in terms of information-theoretic quantities~\cite{Sensory}. Suppressing both work fluctuation and dissipation as much as possible is vital to heat engines and thermodynamic tasks since reducing dissipation allows us to increase the efficiency and reducing work fluctuation makes it possible to supply an exact amount of work needed to complete a given task or to extract a definite amount of work from the system.


Considerable efforts have been devoted in search for a protocol that minimizes work fluctuation and dissipation under nonequilibrium situations. Previous studies have explored the regime around vanishing work fluctuation by using techniques known as single-shot statistical mechanics~\cite{Aberg,Horodecki,Fernando1,Fernando2,Lostaglio}, and the regime around vanishing dissipation on the basis of the second law of thermodynamics~\cite{Hasegawa,Esposito2}. However, as we prove in the present work, these two aims (vanishing work fluctuation and vanishing dissipation) are incompatible. We find the trade-off relation between work fluctuation and dissipation with its fundamental lower bound set by the information distance characterizing the nonequilibriumness of the system. We also show that the bounds on dissipation in the single-shot (vanishing work fluctuation) and reversible (vanishing dissipation) regimes can be smoothly connected via the relative entropy~\cite{Nielsen} and the Renyi divergence~\cite{Renyi}, both of which quantify the information distance between the nonequilibrium distribution and the canonical distribution. We apply the trade-off relation to information heat engines, where the fundamental lower bound of the trade-off relation is characterized by the obtained information. Numerical simulations on an information heat engine based on a single-electron box~\cite{Koski1,Koski2} are performed to verify the trade-off relation. We propose a method to construct explicit protocols that achieve the lower bound of the trade-off relation.

\noindent{\it Main Results.---} We define the extractable work from the system as a change of the internal energy that is not absorbed by the heat bath: $W[\Gamma]=E_{\lambda_{0}}(x)-E_{\lambda_{1}}(y)+ Q[\Gamma]$, where $\Gamma$ denotes the trajectory of the process, $Q[\Gamma]$ is the heat absorbed by the system, and $E_{\lambda_{0}}(x)$ and $E_{\lambda_{1}}(y)$ are the initial and final energies of the eigenstates, respectively. For nonequilibrium initial and final states, the maximum extractable work from the system is quantified by the nonequilibrium free-energy difference~\cite{Esposito2,Parrondo} $-\Delta \mathcal{F}(x,y)=\mathcal{F}_{\lambda_{0}}(x)-\mathcal{F}_{\lambda_{1}}(y)$, where $\mathcal{F}_{\lambda_{0}}(x)=E_{\lambda_{0}}(x)- \beta^{-1}S[p_{\mathrm{ini}}(x)]$, $S[q(x)]=-\ln q(x)$ is the Shannon entropy and $\beta$ is the inverse temperature of the heat bath. We define dissipation as the difference between the maximum extractable work and the actually extracted work:
\beq
\sigma[\Gamma]=-\beta(W[\Gamma]+\Delta \mathcal{F}(x,y)) . \label{noneqentw}
\eeq

\begin{figure*}[tbp]
\begin{center}
\includegraphics[width=\textwidth]{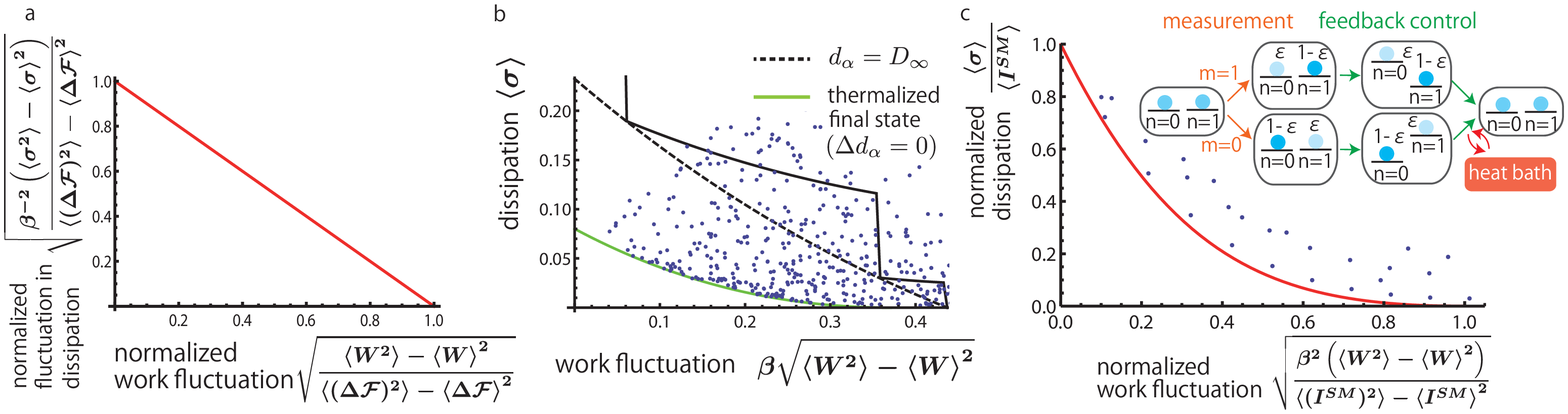}
\caption{ {\bf Trade-off relations.} (a) Normalized standard deviation of dissipation $\sigma$ versus that of work $W$. The solid line shows the lower bound of the trade-off relation~(\ref{de}). (b) Average dissipation versus the standard deviation of work. The black solid curve shows the lower bound of the trade-off relation~(\ref{sigmafluccond}) and (\ref{sigmaflucdistradeoff}) for arbitrary initial and final states. If $d_{\alpha}$ takes the minimum value $D_{\infty}$, the lower bound is given by the dashed curve. For a thermalized final state, $\Delta d_{\alpha}(p_{\mathrm{fin}}||p^{\mathrm{can}}_{\lambda_{1}})=0$ and the lower bound is given by the green solid curve. Each blue dot is obtained by a numerical simulation of a random quench of a five-level system followed by thermalization and isothermal expansion (see Supplementary material for details). (c) The abscissa shows the standard deviation of work normalized by that of the fluctuation of the obtained information, and the ordinate shows the dissipation normalized by the mutual information between the system and the measuring apparatus. The solid curve shows the lower bound of the trade-off relation~(\ref{infoflucdistradeoff}) and (\ref{infoalpha}). Blue dots are obtained by a numerical simulation of a Szilard engine in a single-electron box~\cite{Koski2} as illustrated in the inset (see Supplementary material for details). Here $n$ denotes the excess number of electrons in the quantum dot, $m$ denotes the outcome of the measurement on $n$, and $\epsilon$ is the measurement error rate which is set to be $\epsilon=0.02$ in the numerical simulation. The relevant two states $n=0$ and $n=1$ are assumed to be degenerate and initially populated with equal probability. Depending on the outcome $m$ of the state measurement, the feedback control is performed by lowering the energy level of the $m$ state relative to the other. Finally, the two energy levels are relaxed to its initial (equal-energy) state through thermal contact with a heat bath.}
\label{fig:pidecay}
\end{center}
\end{figure*}

The first main result  of our work is the trade-off relation between work fluctuation and fluctuation in dissipation  (see Supplementary material for the proof):
\beqa
& &\sqrt{\av{W^{2}}-\av{W}^{2}}+\beta^{-1}\sqrt{\av{\sigma^{2}}-\av{\sigma}^{2}} \nonumber \\
& &\hspace{40mm}\geq \sqrt{\av{(\Delta \mathcal{F})^{2}}-\av{\Delta \mathcal{F}}^{2}}. \label{de}
\eeqa
This result implies that the sum of the work fluctuation and the fluctuation in dissipation is bounded from below by the fluctuation of the nonequilibrium free-energy difference $\Delta \mathcal{F}$ (See Fig.~\ref{fig:pidecay} (a)). If the initial and final states are far from equilibrium, the lower bound of~(\ref{de}) becomes very large. The trade-off relation~(\ref{de}) indicates that work and dissipation cannot simultaneously take definite values; if we reduce work fluctuation, the fluctuation in entropy production inevitably increases, and vice versa.

\begin{figure*}[t]
\begin{center}
\includegraphics[width=.9\textwidth]{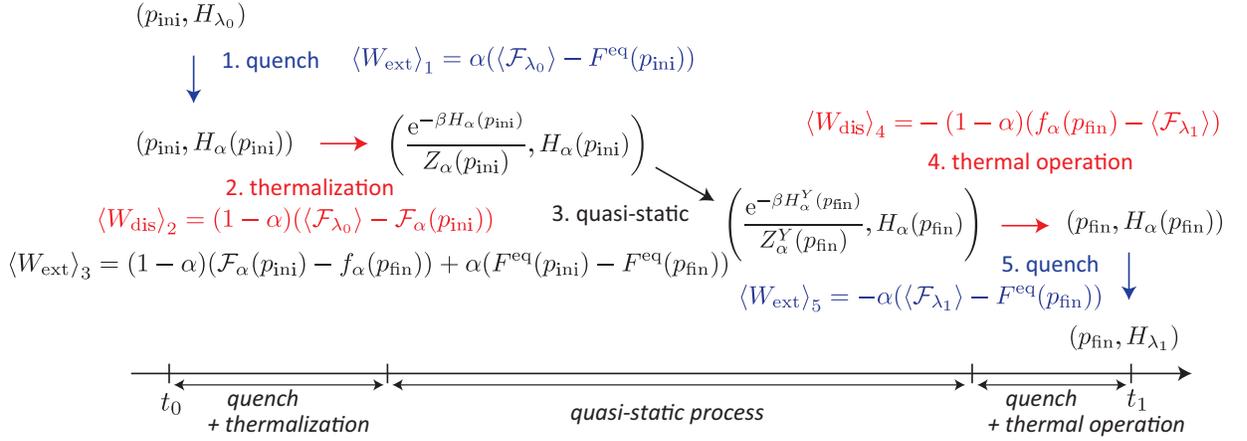}
\caption{ Protocol achieving the lower bounds of the trade-off relations. We denote $(q,H)$ as a pair of the state $q$ and the Hamiltonian $H$. The transformation $(p_{\mathrm{ini}},H_{\lambda_{0}})\rightarrow (p_{\mathrm{fin}},H_{\lambda_{1}})$ that achieves minimum work fluctuation and dissipation is illustrated, where a change in the Hamiltonian is shown in a vertical direction and a change in the state is shown in a horizontal direction. The explicit protocol consists of five steps, where the extractable work $\av{W_{\mathrm{ext}}}$ and the dissipated work $\av{W_{\mathrm{dis}}}:=\beta^{-1}\av{\sigma}$ for each process are shown. Here, $H_{\alpha}(p_{\mathrm{fin}})$ is a Hamiltonian which satisfies $\ee^{-\beta H_{\alpha}(p_{\mathrm{fin}})}/Z_{\alpha}(p_{\mathrm{fin}}):=(p_{\mathrm{fin}})^{\alpha}(p^{\mathrm{can}}_{\lambda_{1}})^{1-\alpha}\ee^{(1-\alpha)D_{\alpha}(p_{\mathrm{fin}}||p^{\mathrm{can}}_{\lambda_{1}})}$.  }
\label{fig:protocol}
\end{center}
\end{figure*} 

The second main result is the trade-off relation between work fluctuation and dissipation. From~(\ref{de}), there is a nontrivial relation between $\sigma$ and $W$ if $\av{W^{2}}-\av{W}^{2}\leq \av{(\Delta \mathcal{F})^{2}}-\av{\Delta \mathcal{F}}^{2}$. Then, let 
\beq
\alpha=\frac{\sqrt{\av{W^2}-\av{W}^{2}}}{ \sqrt{\av{(\Delta \mathcal{F})^{2}}-\av{\Delta \mathcal{F}}^{2}}} , \label{sigmafluccond}
\eeq
where $\alpha\in [0,1]$. In this case, dissipation and work satisfy the following inequalities (see Supplementary material for the proof): 
\beqa
\av{\sigma} &\geq& (1-\alpha)(\Delta \mathcal{D}_{\alpha}(p_{\mathrm{ini}}||p_{\lambda_{0}}^{\mathrm{can}})+\Delta d_{\alpha}(p_{\mathrm{fin}}||p_{\lambda_{1}}^{\mathrm{can}})), \label{sigmaflucdistradeoff} \\ 
\av{W}&\leq&  -\alpha \av{\Delta \mathcal{F}}-(1-\alpha)\left(f_{\alpha}(p_{\mathrm{fin}})- \mathcal{F}_{\alpha}(p_{\mathrm{ini}})\right) .\label{workflucdistradeoff}
\eeqa
Here, $\Delta \mathcal{D}_{\alpha}=D-D_{\alpha}$ ($\Delta d_{\alpha}=d_{\alpha}-D$) gives the distance between the initial (final) distribution and the canonical distribution, $D(p_{\mathrm{ini}}||p_{\lambda_{0}}^{\mathrm{can}})=\sum_{x}p_{\mathrm{ini}}(x)\ln\frac{p_{\mathrm{ini}}(x)}{p_{\lambda_{0}}^{\mathrm{can}}(x)}$ is the Kullback-Leibler divergence (relative entropy)~\cite{Nielsen} and 
\beq
D_{\alpha}(p_{\mathrm{ini}}||p^{\mathrm{can}}_{\lambda_{0}})=\frac{1}{\alpha-1}\ln\biggl[\sum_{x}(p_{\mathrm{ini}}(x))^{\alpha}(p_{\lambda_{0}}^{\mathrm{can}}(x))^{1-\alpha}\biggr] \label{Renyidiv}
\eeq
is the Renyi divergence~\cite{Renyi}. Here $d_{\alpha}$ is defined by
\beq
d_{\alpha}(p_{\mathrm{fin}}||p_{\lambda_{1}}^{\mathrm{can}})=\frac{1}{\alpha-1}\ln\left[\sum_{y\in Y}(p_{\mathrm{fin}}(y))^{\alpha}(p_{\lambda_{1}}^{\mathrm{can}}(y))^{1-\alpha}\right], \label{dalphadiv}
\eeq
where the support $Y$ is defined such that $d_{\alpha}$ takes the smallest value that satisfies 
\beq
d_{\alpha}(p_{\mathrm{fin}}||p_{\lambda_{1}}^{\mathrm{can}})\geq D_{\infty}(p_{\mathrm{fin}}||p_{\lambda_{1}}^{\mathrm{can}})=\ln\max_{y}\frac{p_{\mathrm{fin}}(y)}{p_{\lambda_{1}}^{\mathrm{can}}(y)}. \label{dalphabd} 
\eeq
The lower bound of~(\ref{sigmaflucdistradeoff}) is given by the black solid curve in Fig.~\ref{fig:pidecay} (b). The asymmetry between $D_{\alpha}$ and $d_{\alpha}$ is due to the absence of the time-reversed protocol of the thermalization process as discussed later. In~(\ref{workflucdistradeoff}), $\av{\mathcal{F}_{\lambda_{0}}}=\beta^{-1}D(p_{\mathrm{ini}}||p_{\lambda_{0}}^{\mathrm{can}})+F^{\mathrm{eq}}(p^{\mathrm{can}}_{\lambda_{0}})$  is the averaged nonequilibrium free-energy, and $\mathcal{F}_{\alpha}(p_{\mathrm{ini}})=\beta^{-1}D_{\alpha}(p_{\mathrm{ini}}||p^{\mathrm{can}}_{\lambda_{0}})+F^{\mathrm{eq}}(p^{\mathrm{can}}_{\lambda_{0}})$ is the $\alpha$-generalization of the free energy, where we denote $F^{\mathrm{eq}}(q)$ as the equilibrium free energy whose corresponding canonical distribution is equal to the distribution $q$. We also define the free energy $ f_{\alpha}(p_{\mathrm{fin}})=\beta^{-1}d_{\alpha}(p_{\mathrm{fin}}||p_{\lambda_{1}}^{\mathrm{can}})+ F^{\mathrm{eq}}(p^{\mathrm{can}}_{\lambda_{1}})$  by using $d_{\alpha}$. We note that the ordering of the Renyi divergence~\cite{Ervin1} $D_{\infty}\geq D\geq D_{\alpha}$ for $1\geq\alpha$ with~(\ref{dalphabd}) implies $\Delta D_{\alpha}\geq 0$ and $\Delta d_{\alpha}\geq 0$.

\noindent{\it Explicit protocol and the trade-off relation.---} For given $(p_{\mathrm{ini}},H_{\lambda_{0}})$ and $(p_{\mathrm{fin}},H_{\lambda_{1}})$, we want to find a protocol which connects them by reducing both work fluctuation and dissipation as much as possible. Although a quasi-static process makes both work fluctuation and dissipation vanish, we cannot directly connect $(p_{\mathrm{ini}},H_{\lambda_{0}})\rightarrow(p_{\mathrm{fin}},H_{\lambda_{1}})$ by the quasi-static process alone because the initial and final distributions are out of equilibrium. Instead, we prepare two canonical distributions as auxiliary intermediate states, and connect them by the quasi-static process. Then, we connect $(p_{\mathrm{ini}},H_{\lambda_{0}})$ with one of the canonical distributions by combining a quench process followed by thermalization, and the other canonical distribution is connected with $(p_{\mathrm{fin}},H_{\lambda_{1}})$ by a thermal operation and a quench process. The entire protocol is illustrated in Fig.~\ref{fig:protocol}, and as we show in the Supplementary material, this protocol is necessary and sufficient to achieve the lower bound of~(\ref{de}) and (\ref{sigmaflucdistradeoff}).

Now let us discuss the explicit protocol in more detail and consider the physical meanings of the quantities that appear in~(\ref{sigmaflucdistradeoff}) and (\ref{workflucdistradeoff}). We change the initial distribution to the canonical distribution $\ee^{-\beta H_{\alpha}(p_{\mathrm{ini}})}/Z_{\alpha}(p_{\mathrm{ini}}):=(p_{\mathrm{ini}})^{\alpha}(p^{\mathrm{can}}_{\lambda_{0}})^{1-\alpha}\ee^{(1-\alpha)D_{\alpha}(p_{\mathrm{ini}}||p^{\mathrm{can}}_{\lambda_{0}})}$, which is an intermediate distribution between the initial state and the canonical distribution for the initial Hamiltonian. This is done by quenching the Hamiltonian from $H_{\lambda_{0}}$ to $H_{\alpha}(p_{\mathrm{ini}})$ and extract the work given by $\av{W_{\mathrm{ext}}}_{1}=\alpha(\av{\mathcal{F}_{\lambda_{0}}}-F^{\mathrm{eq}}(p_{\mathrm{ini}}))$. Note that the maximum extractable from from the initial state is quantified by the nonequilibrium free energy $\av{\mathcal{F}_{\lambda_{0}}}$. The unexpended free energy $(1-\alpha)\av{\mathcal{F}_{\lambda_{0}}}$ is partly lost during the thermalization, and the remaining free energy, which can be extracted by the quasi-static process, is given by $(1-\alpha)\mathcal{F}_{\alpha}(p_{\mathrm{ini}})$, as can be seen by noting the dissipated work due to the measurement: $\av{W_{\mathrm{dis}}}_{2}=(1-\alpha)(\av{\mathcal{F}_{\lambda_{0}}}-\mathcal{F}_{\alpha}(p_{\mathrm{ini}}))$. This dissipation $\beta\av{W_{\mathrm{dis}}}_{2}=(1-\alpha)\Delta D_{\alpha}(p_{\mathrm{ini}}||p^{\mathrm{can}}_{\lambda_{0}})=D(p_{\mathrm{ini}}||\ee^{-\beta H_{\alpha}(p_{\mathrm{ini}})}/Z_{\alpha}(p_{\mathrm{ini}}))$ appears on the right-hand side of~(\ref{sigmaflucdistradeoff}), which gives the information distance between the initial state and the canonical distribution which we connect during the thermalization process. Thus, the right-hand side of~(\ref{workflucdistradeoff}) is comprised of a part of the nonequilibrium free energy $\alpha\av{\mathcal{F}_{\lambda_{0}}}$ which can be extracted by the quench process and the free energy $(1-\alpha)\mathcal{F}_{\alpha}(p_{\mathrm{ini}})$ which remains in the system after the thermalization.

The rest of the protocol is the transformation of the canonical distribution to the final state. Because we cannot perform time-reversal of the thermalization, we invoke a thermal operation~\cite{Horodecki} which transforms the state of the system by exchanging energy with the heat bath. This operation always changes the system closer to the thermal equilibrium, and we need to prepare a distribution whose ``nonequilibriumness" is larger than that of the target final state. For this purpose, we prepare a localized distribution $\ee^{-\beta H_{\alpha}^{Y}(p_{\mathrm{fin}})}/Z^{Y}_{\alpha}(p_{\mathrm{fin}}):=(p_{\mathrm{fin}})^{\alpha}(p^{\mathrm{can}}_{\lambda_{1}})^{1-\alpha}\ee^{(1-\alpha)d_{\alpha}(p_{\mathrm{fin}}||p^{\mathrm{can}}_{\lambda_{1}})}$, whose support is restricted to $Y$. The term $(1-\alpha)f_{\alpha}(p_{\mathrm{fin}})$ is the free energy which is needed to prepare this localized thermal state and $\alpha\av{\mathcal{F}_{\lambda_{1}}}$ is the free energy needed to quench the Hamiltonian back to the final one (see (\ref{workflucdistradeoff}) and Fig.~\ref{fig:protocol}). The asymmetry between the transformation of a nonequilibrium state into a thermalized state and its opposite transformation (i.e., from a thermalized state to a nonequilibrium state) gives rise to the difference between $D_{\alpha}$ and $d_{\alpha}$ (see~(\ref{sigmaflucdistradeoff})).

As shown in Ref.~\cite{Horodecki}, the minimum work cost to create $p_{\mathrm{fin}}$ from a canonical distribution via the thermal operation with the fixed Hamiltonian $H_{\alpha}(p_{\mathrm{fin}})$ is given by $\beta^{-1}D_{\infty}(p_{\mathrm{fin}}||\ee^{-\beta H_{\alpha}(p_{\mathrm{fin}})}/Z_{\alpha}(p_{\mathrm{fin}}))$, with the help of a two-level auxiliary system. If we can introduce this auxiliary system, the dissipated work for the thermal operation is found to be $\av{W_{\mathrm{dis}}}_{4}=(1-\alpha)(\mathcal{F}_{\infty}(p_{\mathrm{fin}})-\av{\mathcal{F}_{\lambda_{1}}})$, and the equality condition in~(\ref{dalphabd}) is achieved (see Supplementary material for details). This condition is also achieved if the energy level of the system is dense. The lower bound of~(\ref{sigmaflucdistradeoff}) with $d_{\alpha}=D_{\infty}$ is shown by the dashed curve in Fig.~\ref{fig:pidecay} (b). Note that the solid curve jumps (i.e., the support $Y$ changes) wherever the line touches the dashed curve because we take discrete energy levels.

\noindent{\it Comparison with previous studies.---} For $\alpha=1$, (\ref{sigmaflucdistradeoff}) and (\ref{workflucdistradeoff}) are equivalent to the second law of thermodynamics for arbitrary initial and final states: $\av{\sigma}\geq 0$ and $\av{W}\leq -\av{\Delta \mathcal{F}}$. Since the canonical distribution $\ee^{-\beta H_{\alpha}(p_{\mathrm{ini}})}/Z_{\alpha}(p_{\mathrm{ini}})$ is equal to the initial state for $\alpha=1$ (the same relation holds for the final state), we do not need thermalization and the thermal operation to achieve the lower bound of the trade-off relations. Then, dissipation does not occur and we can extract the maximum average work from the system (see also Fig.~\ref{fig:protocol}).  For $\alpha=0$, (\ref{workflucdistradeoff}) takes the form $\av{W}\leq \mathcal{F}_{0}(p_{\mathrm{ini}})-f_{0}(p_{\mathrm{fin}})$, which reproduce the single-shot results given in Refs.~\cite{Aberg,Horodecki}. Here, $\mathcal{F}_{0}(p_{\mathrm{ini}})=-\beta^{-1}\ln [\sum_{x\in X}\exp(-\beta E_{\lambda_{0}}(x))]$ is equal to the equilibrium local free energy whose support $X$ is the same as the initial state. By raising the initially unoccupied energy levels, this amount of free energy remains after the thermalization.

For a general $\alpha$, the trade-off relation gives the minimum amount of work fluctuation and dissipation in the intermediate regime. Comparing (\ref{sigmafluccond}) and~(\ref{workflucdistradeoff}), we find that the distribution of the extractable work is broadened (meaning larger work fluctuation) if we want to increase the average value of work, and vice versa. Thus, the trade-off relation gives the best combinations of the ``quality of work" and the average amount of extractable work. For equilibrium initial and final states, we can directly connect them by the quasi-static process and the lower bound of the trade-off relation (solid curve) in Fig.~\ref{fig:pidecay} (b) shrinks to a single point at the origin, i.e., work fluctuation and dissipation can both vanish.

\noindent{\it Applications to information heat engines.---} The information heat engines utilize the information obtained by the measurement to extract work from the system.  For simplicity, we consider a classical system and assume that the premeasurement state $p^{S}(x)$ is given by a canonical distribution. Then, dissipation is defined as the difference between the maximum amount of extractable work~\cite{Sagawa1} $-\Delta F^{S}+\beta^{-1}I^{SM}$ and the actually extracted work $W^{S}[\Gamma]$:
\beq
\sigma[\Gamma]=-\beta(W^{S}[\Gamma]+\Delta F^{S})+I^{SM}(x,a),
\eeq
and $I^{SM}(x,a)=\ln p^{SM}(x,a)-\ln(p^{S}(x)p^{M}(a))$ is the (unaveraged) classical mutual information between the system ($S$) and the measurement apparatus ($M$)~\cite{Nielsen}. Here, $p^{SM}(x,a)$ is the joint probability distribution of $SM$ for the postmeasurement state,  $p^{S}(x)=\sum_{a}p^{SM}(x,a)$ and $p^{M}(a)=\sum_{x}p^{SM}(x,a)$. 
The trade-off relation~(\ref{sigmaflucdistradeoff}) takes the following form (see also Fig.~\ref{fig:pidecay} (c)):
\beqa
\av{\sigma}&\geq& (1-\alpha)(I^{SM}-I^{SM}_{\alpha}), \label{infoflucdistradeoff} \\ 
\beta\av{W}&\leq&  \alpha \av{I^{SM}}+(1-\alpha)I^{SM}_{\alpha}-\beta\Delta F^{S}  ,\label{infoworkflucdistradeoff}
\eeqa
where $\alpha$ is defined by
\beq
\alpha=\frac{ \beta \sqrt{ \av{W^2}-\av{W}^{2}} }{ \sqrt{\av{(I^{SM})^{2}}-\av{I^{SM}}^{2}} }, \label{infoalpha}
\eeq
and $I^{SM}_{\alpha} =\frac{1}{\alpha-1}\ln\sum_{x,a}(p^{SM}(x,a))^{\alpha}(p^{S}(x)p^{M}(a))^{1-\alpha}$ is the Renyi generalization of the mutual information. If we extract the maximum amount of work from the system for each measurement outcome, we can extract $W^{S}[\Gamma]=-\Delta F^{S}+\beta^{-1}I(x,a)$ from the system, with finite work fluctuations. On the other hand, if we discard the measurement outcome, we can extract a definite amount of work $W^{S}[\Gamma]=-\Delta F^{S}$ from the system with large dissipation. This means that~(\ref{infoflucdistradeoff}) and~(\ref{infoalpha}) show a trade-off relation between work fluctuation and dissipation due to the fluctuation in the obtained information. 

\noindent{\it Possible experimental test of the trade-off relations.---} The proposed trade-off relations can be tested by using the single electron box, which was used to realize a Szilard engine~\cite{Koski1,Koski2}. Suppose that we prepare degenerate states of a two-level system and perform measurement to distinguish the state of the system which is initially distributed with equal probabilities $P(n=0)=P(n=1)=1/2$, where $n$ labels the state of the system. Let the measurement error rate be $\epsilon$ and the joint probability distribution of the system being $n$ and the measurement outcome being $m$ be given by $P(n,m)=(1-\epsilon)/2$ for $m=n$ and $P(n,m)=\epsilon/2$ for $m\neq n$. A feedback control is implemented by lowering the energy level of the state $n=m$ and let the energy-level return to the degeneracy point (see the inset of Fig.~\ref{fig:pidecay} (c)). By tracking the state of the system during this feedback, we can measure the extracted work for each run of the experiment, and calculate work fluctuation and dissipation. If we change the feedback protocol, e.g., by changing the degree of the energy-level shift, we obtain a different experimental data set of work fluctuation and dissipation. By plotting $\beta\sqrt{ \av{W^2}-\av{W}^{2}} / \sqrt{\av{(I^{SM})^{2}}-\av{I^{SM}}^{2}}$ against $\av{\sigma}/\av{I^{SM}}$ as shown in Fig.~\ref{fig:pidecay} (c), we can test the trade-off relation between work fluctuation and dissipation in information heat engines. The results of numerical simulations of a Szilard engine in a single-electron box using a master equation described in Ref.~\cite{Koski2} are shown as dots in Fig.~\ref{fig:pidecay} (c).


\noindent{\it Summary.---} We have found a set of fundamental trade-off relations between work fluctuation and dissipation for nonequilibrium initial and final states. We can reproduce single-shot results in the limit of vanishing work fluctuation and thermodynamically reversible results (the lower bound of the conventional second law) in the limit of vanishing dissipation. These two limits are smoothly connected and the minimum dissipation along this boundary is characterized by the information distance between the state of the system and the canonical distribution. This result gives the fundamental bound on both work fluctuation and dissipation starting from and/or ending at nonequilibrium states. An application of the trade-off relation to information heat engines is discussed, including numerical simulations which vindicate the obtained trade-off relation.

\begin{acknowledgments}
This work was supported by KAKENHI Grant No. 26287088 from the Japan Society for the Promotion of Science, a Grant-in-Aid for Scientific Research on Innovative Areas `Topological Materials Science' (KAKENHI Grant No. 15H05855), the Photon Frontier Network Program from MEXT of Japan, and the Mitsubishi Foundation. K. F. acknowledges support from JSPS (Grant No. 254105) and through Advanced Leading Graduate Course for Photon Science (ALPS). K. F. thanks Hal Tasaki, Takahiro Sagawa, Jukka Pekola, Jonne Koski, Y\^{u}to Murashita, Tomohiro Shitara, Yusuke Horinouchi and Kohaku So for fruitful discussions and comments.

\end{acknowledgments}

\onecolumngrid
\appendix

\section{Proof of the first main result}
To show the first main result in Eq.~(2) of the main text
\beq
\sqrt{\av{W^{2}}-\av{W}^{2}}+\beta^{-1}\sqrt{\av{\sigma^{2}}-\av{\sigma}^{2}}\geq \sqrt{\av{(\Delta \mathcal{F})^{2}}-\av{\Delta \mathcal{F}}^{2}},  \label{supde}
\eeq
we first calculate the variance of $\sigma+\beta W$. By using Eq.~(1) in the main text
\beq
\sigma[\Gamma]=-\beta(W[\Gamma]+\Delta \mathcal{F}(x,y)) , \label{sup:noneqentw}
\eeq
we obtain
\beq
\beta^{2}(\av{W^{2}}-\av{W}^{2}) +\av{\sigma^{2}}-\av{\sigma}^{2}+2\beta(\av{\sigma W}-\av{\sigma}\av{W}) =\av{(\Delta \mathcal{F})^{2}}-\av{\Delta \mathcal{F}}^{2} . \label{da}
\eeq
Next, we consider the property of the variance-covariance matrix defined by
\beq
V_{ij}=\av{(X_{i}-\av{X_{i}})(X_{j}-\av{X_{j}})}.
\eeq
The eigenvalues of $V_{ij}$ are positive semi-definite and thus the determinant of $V_{ij}$ is nonnegative. By taking $X_{1}=\beta W$ and $X_{2}=\sigma$, and from $\text{Det}[V_{ij}]\geq 0$, we obtain
\beq
(\av{W^{2}}-\av{W}^{2})(\av{\sigma^{2}}-\av{\sigma}^{2})\geq (\av{\sigma W}-\av{\sigma}\av{W})^{2} . \label{db}
\eeq
Taking the square root of~(\ref{db}), we obtain 
\beq
\sqrt{\av{W^{2}}-\av{W}^{2}}\sqrt{\av{\sigma^{2}}-\av{\sigma}^{2}}\geq \av{\sigma W}-\av{\sigma}\av{W}. \label{ddb}
\eeq
Note that the above inequality holds trivially for $\av{\sigma W}-\av{\sigma}\av{W}\leq 0$. Combining~(\ref{da}) and~(\ref{ddb}), we obtain
\beq
\left(\sqrt{\av{W^{2}}-\av{W}^{2}}+\beta^{-1}\sqrt{\av{\sigma^{2}}-\av{\sigma}^{2}}\right)^{2}\geq \av{(\Delta \mathcal{F})^{2}}-\av{\Delta \mathcal{F}}^{2} . \label{dc}
\eeq
By taking the square root of either side of~(\ref{dc}), we obtain the trade-off relation~(\ref{supde}).

The equality condition in~(\ref{supde}) is satisfied if and only if one of the eigenvalues of the matrix $V_{ij}$ is zero, i.e., if and only if there exist some constants $a$ and $b$ such that the variance of $a\sigma-b\beta W$ vanishes. Without the loss of generality, we can take $a\geq 0$.

\section{Proof of the second main result for an equilibrium final state}
We first assume $p_{\mathrm{fin}}(y)=p_{\lambda_{1}}^{\mathrm{can}}(y)$ and prove the second main result. We discuss a more general case in Sec.~\ref{sec:noneqfinal}.  From the assumption, Eq.~(\ref{sup:noneqentw}) is given by
\beqa
\sigma[\Gamma] &=&-\beta W[\Gamma] -\beta F^{\mathrm{eq}}(p^{\mathrm{can}}_{\lambda_{1}}) +  \mathcal{F}_{\lambda_{0}}(x) \nonumber \\
&=&-\beta W[\Gamma]-\beta \Delta F^{\mathrm{eq}} +\ln \frac{p_{\mathrm{ini}}(x)}{p_{\lambda_{0}}^{\mathrm{can}}(x)}, \label{sup:sigw}
\eeqa
where $\Delta F^{\mathrm{eq}}=F^{\mathrm{eq}}(p^{\mathrm{can}}_{\lambda_{1}})-F^{\mathrm{eq}}(p^{\mathrm{can}}_{\lambda_{0}})$ is the equilibrium free-energy difference, and the trade-off relation~(\ref{supde}) takes the form
\beq
\sqrt{\av{W^{2}}-\av{W}^{2}}+\beta^{-1}\sqrt{\av{\sigma^{2}}-\av{\sigma}^{2}}\geq \sqrt{\av{( \mathcal{F}_{\lambda_{0}})^{2}}-\av{\mathcal{F}_{\lambda_{0}}^{2}}}. \label{deee}
\eeq

\subsection{Detailed fluctuation theorem} 
We use the detailed fluctuation theorem which relates the ratio of the path probabilities and the total entropy production to prove the main results:
\beq
\frac{P[\Gamma]}{\tilde{P}[\Gamma^{\dagger}]}=e^{\sigma[\Gamma]}, \label{df}
\eeq
where $\Gamma$ and $\Gamma^{\dagger}$ denote the trajectories of the forward and backward (time-reversed) processes, and $P[\Gamma]$ and $\tilde{P}[\Gamma^{\dagger}]$ are the corresponding path probability distributions. Here,
\beq
\sigma[\Gamma]=\Delta s[\Gamma]-\beta Q[\Gamma] \label{totent}
\eeq
is the total entropy production, where $\Delta s[\Gamma]=\ln p_{\mathrm{ini}}(x)-\ln p_{\mathrm{fin}}(y)$ is a change in the Shannon entropy of the system. By using the definition of the nonequilibrium free energy, the total entropy production~(\ref{totent}) is equal to Eq.~(\ref{sup:noneqentw}). Note that Eq.~(\ref{df}) can be derived classically~[26,27] and quantum-mechanically~[27-30] for general settings (e.g., in the Hamiltonian dynamics and the stochastic dynamics). The following argument can also be applied to quantum systems if the initial density matrix of the system is diagonal in the initial energy eigenbasis. 

\subsection{Equality condition of the inequality~(\ref{deee})}
Combining the detailed fluctuation theorem~(\ref{df}) and Eq.~(\ref{sup:sigw}), we have
\beq
a\sigma[\Gamma]-b\beta W[\Gamma]=b\beta\Delta F^{\mathrm{eq}}+(a+b)\ln\frac{P[\Gamma]}{\tilde{P}[\Gamma^{\dagger}]}-b\ln\frac{p_{\mathrm{ini}}(x)}{p_{\lambda_{0}}^{\mathrm{can}}(x)}. \label{dff}
\eeq
We set the condition
\beq
a\sigma[\Gamma]-b\beta W[\Gamma]=b\beta\Delta F^{\mathrm{eq}}+c \label{dg}
\eeq
to make the variance of $a\sigma-b\beta W$ vanish, where $c$ is a constant. Then, Eqs.~(\ref{dff}) and (\ref{dg}) lead to
\beq
\frac{P[\Gamma]}{\tilde{P}[\Gamma^{\dagger}]}\left(\frac{p_{\mathrm{ini}}(x)}{p_{\lambda_{0}}^{\mathrm{can}}(x)}\right)^{\frac{-b}{a+b}}=\exp(\frac{c}{a+b}). \label{dj}
\eeq
The normalization of the backward probability distribution, i.e., $\sum_{\Gamma}\tilde{P}[\Gamma^{\dagger}]=1$, combined with the relation $P[\Gamma]=P[\Gamma|x]p_{\mathrm{ini}}(x)$, fixes the constant $c$:
\beq
\sum_{x}(p_{\lambda_{0}}^{\mathrm{can}}(x))^{\frac{b}{a+b}}(p_{\mathrm{ini}}(x))^{\frac{a}{a+b}}=\exp(\frac{c}{a+b}), \label{dh}
\eeq
where $P[\Gamma|x]$ is the conditional forward probability conditioned on $x$ and $\sum_{\Gamma/x}P[\Gamma|x]=1$. By introducing the Renyi divergence, the constant $c$ takes the form
\beq
\frac{c}{a+b}=-(1-\alpha)D_{\alpha}(p_{\mathrm{ini}}||p_{\lambda_{0}}^{\mathrm{can}}),
\eeq
where $\alpha=a/(a+b)$. Using Eq.~(\ref{dh}), $\sigma$ and $W$ take the following forms:
\beqa
\sigma[\Gamma]&=&(1-\alpha)\left(\mathcal{D}_{\lambda_{0}}(x)-D_{\alpha}(p_{\mathrm{ini}}||p_{\lambda_{0}}^{\mathrm{can}})\right), \label{bdsigmaun} \\
\beta W[\Gamma]&=&-\beta\Delta F^{\mathrm{eq}}+\alpha \mathcal{D}_{\lambda_{0}}(x) +(1-\alpha)D_{\alpha}(p_{\mathrm{ini}}||p_{\lambda_{0}}^{\mathrm{can}}) , \label{bdwun}
\eeqa
where $\mathcal{D}_{\lambda_{0}}(x)=\ln p_{\mathrm{ini}}(x) - \ln p_{\lambda_{0}}^{\mathrm{can}}(x)$. Note that $\mathcal{F}_{\lambda_{0}}(x)=F^{\mathrm{eq}}(p^{\mathrm{can}}_{\lambda_{0}})+\mathcal{D}_{\lambda_{0}}(x)$. In particular, the first and the second terms on the left-hand side of~(\ref{deee}) are given by
\beqa
\sqrt{\av{\sigma^{2}}-\av{\sigma}^{2}}&=&\beta^{-1}|1-\alpha| \sqrt{\av{(\mathcal{F}_{\lambda_{0}})^{2}}-\av{\mathcal{F}_{\lambda_{0}}}^{2}}, \label{boundarysigma} \\
\sqrt{\av{W^{2}}-\av{W}^{2}}&=& \alpha \sqrt{\av{(\mathcal{F}_{\lambda_{0}})^{2}}-\av{\mathcal{F}_{\lambda_{0}}}^{2}}.
\eeqa
We conclude that the equality condition in~(\ref{deee}) is satisfied if and only if Eqs.~(\ref{bdsigmaun}) and (\ref{bdwun}) hold and $\alpha$ takes the value in a range $0\leq \alpha\leq 1$.  

\subsection{Proof of the second main result}

Let us derive the second main result ((4) in the main text) for the case of $p_{\mathrm{fin}}=p_{\lambda_{1}}^{\mathrm{can}}$:
\beq
\av{\sigma} \geq (1-\alpha)( D(p_{\mathrm{ini}}||p_{\lambda_{0}}^{\mathrm{can}}) - \mathcal{D}_{\alpha}(p_{\mathrm{ini}}||p_{\lambda_{0}}^{\mathrm{can}}) ), \label{sup:sigmaflucdistradeoff}
\eeq
where the work fluctuation is assumed to be fixed (Eq.~(3) in the main text) :
\beq
\sqrt{\av{W^2}-\av{W}^{2}}=\alpha\sqrt{\av{(\mathcal{F}_{\lambda_{0}})^{2}}-\av{\mathcal{F}_{\lambda_{0}}}^{2}}  \label{sup:sigmafluccond}
\eeq
where $\alpha\in [0,1]$ is a parameter. We start by combining~(\ref{deee}) and (\ref{sup:sigmafluccond}) and obtain
\beq
\sqrt{ \av{\sigma^2}-\av{\sigma}^{2}} \geq (1-\alpha) \sqrt{\av{\mathcal{D}^{2}}-\av{\mathcal{D}}^{2}}. \label{workflucgeq}
\eeq
Next, we expand the integral fluctuation theorem
\beq
\av{\ee^{-\sigma}}=\sum_{\Gamma}P[\Gamma]\frac{P^{\dagger}[\Gamma^{\dagger}]}{P[\Gamma]}=1 \label{intfluc}
\eeq
around $\sigma[\Gamma]=\sigma_{\alpha}[\Gamma]$ that achieves the lower bound of~(\ref{workflucgeq}):
\beq
\sigma_{\alpha}[\Gamma]=(1-\alpha)\left(\mathcal{D}_{\lambda_{0}}(x)-D_{\alpha}(p_{\mathrm{ini}}||p_{\lambda_{0}}^{\mathrm{can}})\right).
\eeq
From Eq.~(\ref{intfluc}), we obtain
\beqa
1&=&\av{\ee^{-\sigma}}=\sum_{\Gamma}P[\Gamma] \ee^{-(\sigma[\Gamma]-\sigma_{\alpha}[\Gamma])-\sigma_{\alpha}[\Gamma]} \nonumber \\
&=&\sum_{\Gamma}P_{\alpha}[\Gamma] \ee^{-(\sigma[\Gamma]-\sigma_{\alpha}[\Gamma])}. \label{intflucalpha}
\eeqa
Here,
\beq
P_{\alpha}[\Gamma]=P[\Gamma]\left(\frac{p_{\lambda_{0}}^{\mathrm{can}}(x)}{p_{\mathrm{ini}}(x)}\right)^{1-\alpha}\ee^{(1-\alpha)D_{\alpha}(p_{\mathrm{ini}}||p_{\lambda_{0}}^{\mathrm{can}})} 
\eeq
gives the forward probability distribution in which the entropy production is close to $\sigma_{\alpha}[\Gamma]$. In fact, $P_{\alpha}[\Gamma]=\tilde{P}[\Gamma^{\dagger}]$ indicates that $\sigma[\Gamma]=\sigma_{\alpha}[\Gamma]$. Expanding the exponent in Eq.~(\ref{intflucalpha}) for small $\sigma-\sigma_{\alpha}$, we have
\beqa
0&=&\ln\av{\ee^{-(\sigma-\sigma_{\alpha})}}_{\alpha} \nonumber \\
&=&\ln\left\langle 1\right. -(\sigma-\sigma_{\alpha})+\half(\sigma-\sigma_{\alpha})^{2}\left. \right\rangle_{\alpha}+O\left((\sigma-\sigma_{\alpha})^{3}\right).
\eeqa
We thus obtain the fluctuation-dissipation theorem near the point $\sigma_{\alpha}$:
\beq
2\av{\sigma-\sigma_{\alpha}}_{\alpha}=\av{(\sigma-\sigma_{\alpha})^{2}}_{\alpha}-\av{\sigma-\sigma_{\alpha}}_{\alpha}^{2}+O\left((\sigma-\sigma_{\alpha})^{3}\right). \label{flucdissalpha}
\eeq
Since the variance of $\sigma$ is minimized if and only if $\sigma[\Gamma]=\sigma_{\alpha}[\Gamma]$, the average value of $\sigma$ is also minimized for the same condition by using Eq.~(\ref{flucdissalpha}). Combining~(\ref{workflucgeq}) and Eq.~(\ref{flucdissalpha}), we obtain the inequality~(\ref{sup:sigmaflucdistradeoff}).

\section{Explicit protocols that achieve the lower bound of the trade-off relations}
\begin{figure*}[tbp]
\begin{center}
\includegraphics[width=\textwidth]{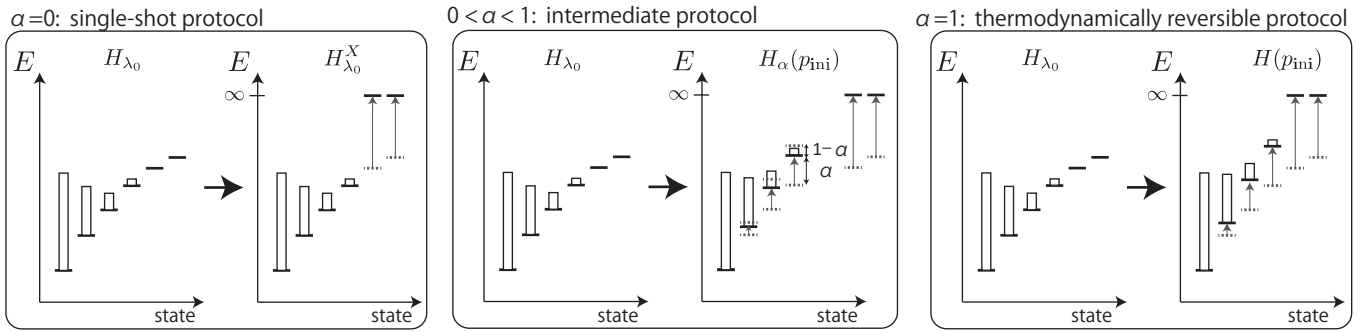}
\caption{ Protocols achieving the lower bounds of the trade-off relations for thermalized final states. The abscissa and ordinate show the state $x$ and energy of the system, respectively. The horizontal bars show the energy levels (the spectrum) of the Hamiltonian and the height of each rectangular box shows the probability distribution of the initial state $p_{\mathrm{ini}}(x)$. We illustrate how the spectrum changes according to each protocol. After the change of the spectrum is completed, we attach a heat bath and quasi-statically change the Hamiltonian of the system to $H_{\lambda_{1}}$. For $\alpha=1$, we change the Hamiltonian to $H(p_{\mathrm{ini}})$, which is defined such that the canonical distribution with respect to $H(p_{\mathrm{ini}})$ is equal to the initial distribution. Dissipation does not occur while the system interacts with a heat bath. For $\alpha=0$, we change the Hamiltonian by raising the energy levels to infinity, whose populations of the initial state are empty~[17]. Note that work fluctuation vanishes during this process. For  general $\alpha$, we change the Hamiltonian to $H_{\alpha}(p_{\mathrm{ini}})=(1-\alpha)H_{\lambda_{0}}+\alpha H(p_{\mathrm{ini}})$, i.e., we mix the changes of the energy levels of two protocols $\alpha=0$ and $\alpha=1$ by the ratio $\alpha:1-\alpha$.}
\label{fig:protocol}
\end{center}
\end{figure*} 
\subsection{Reversible regime ($\alpha=1$)}
Let us first consider the case of $\alpha=1$. The condition~(\ref{dj}) is given by 
\beq
P[\Gamma]=\tilde{P}[\Gamma^{\dagger}] , \label{thermorever}
\eeq
which means that the process is thermodynamically reversible, i.e., the backward protocol exactly brings the final state back into the initial state. The explicit protocol consists of two steps: (1) Change the Hamiltonian from $H_{\lambda_{0}}$ to $H(p_{\mathrm{ini}})$ while keeping the probability distribution $p_{\mathrm{ini}}(x)$ fixed. The new Hamiltonian $H(p_{\mathrm{ini}})$ is chosen such that the canonical distribution with respect to $H(p_{\mathrm{ini}})$ is equal to $p_{\mathrm{ini}}(x)$. (2) We change the Hamiltonian from $H(p_{\mathrm{ini}})$ to $H_{\lambda_{1}}$ slowly (quasi-static isothermal process). Since the time-reversal of these protocols (1) and (2) brings the final state $p_{\lambda_{1}}^{\mathrm{can}}(y)$ back to the initial state $p_{\mathrm{ini}}(x)$, the condition~(\ref{thermorever}) is satisfied. During the protocol (1), we change the energy levels of the system, which leads to a nonzero work fluctuation.
\subsection{Single-shot regime ($\alpha=0$)}
Next, let us consider the case of $\alpha=0$. The condition~(\ref{dj}) is given by 
\beqa
\tilde{P}[\Gamma^{\dagger}]&=&P[\Gamma]\frac{p_{\lambda_{0}}^{\mathrm{can}}(x)}{p_{\mathrm{ini}}(x)}\ee^{D_{0}(p_{\mathrm{ini}}||p_{\lambda_{0}}^{\mathrm{can}})} \nonumber \\
&=& P[\Gamma|x]p_{\lambda_{0}}^{\mathrm{can}, X}(x), \label{thermosingle}
\eeqa
where we introduce a conditional forward probability distribution $P[\Gamma|x]=P[\Gamma]/p_{\mathrm{ini}}(x)$, and a local canonical distribution $p_{\lambda_{0}}^{\mathrm{can}, X}(x)=\exp(-\beta E_{\lambda_{0}}(x) +\beta F_{\lambda_{0}}+D_{0}(p_{\mathrm{ini}}||p_{\lambda_{0}}^{\mathrm{can}}))$ which has the same support $X=\{x|p_{\mathrm{ini}}(x)\neq 0\}$ as the initial distribution. Now the condition~(\ref{thermosingle}) means that we should construct a process starting from $p_{\lambda_{0}}^{\mathrm{can}, X}(x)$ which is thermodynamically reversible. Since the protocol $\lambda_{t}$ starts from a Hamiltonian $H_{\lambda_{0}}$, we need to consider the following two steps: (1) Change the Hamiltonian from $H_{\lambda_{0}}$ to a local Hamiltonian $H^{X}_{\lambda_{0}}=\sum_{x\in X}E_{\lambda_{0}}(x)$ by changing the energy levels labeled by $x\not\in X$ to infinity in the initial Hamiltonian $H_{0}$. Then, $p_{\lambda_{0}}^{\mathrm{can}, X}(x)$ is equal to the canonical distribution with respect to $H^{X}_{\lambda_{0}}$. (2) Change the Hamiltonian from $H^{X}_{\lambda_{0}}$ to $H_{\lambda_{1}}$ slowly in contact with the heat bath.  Since the time reversal of these protocols (1) and (2) brings the final state $p_{\lambda_{1}}^{\mathrm{can}}(y)$ back to $p_{\lambda_{0}}^{\mathrm{can}, X}(x)$, the condition in Eq.~(\ref{thermosingle}) is satisfied. 

Now let us consider how the initial distribution changes by applying the protocols (1) and (2) described above. During the protocol (1), the probability distribution does not change. However, at the beginning of the protocol (2), the initial distribution thermalizes to $p_{\lambda_{0}}^{\mathrm{can}, X}(x)$ because the change of the Hamiltonian from $H^{X}_{\lambda_{0}}$ to $H_{\lambda_{1}}$ is slow. Finally, the distribution of the system is given by $p_{\lambda_{1}}^{\mathrm{can}}(y)$. Note that during the thermalization process, the work fluctuation is zero but the dissipation is nonzero.
\subsection{Intermediate regime (general $\alpha$)}
Finally, we consider the case of general $\alpha$. The condition~(\ref{dj}) is given by 
\beqa
\tilde{P}[\Gamma^{\dagger}]&=&P[\Gamma]\left(\frac{p_{\lambda_{0}}^{\mathrm{can}}(x)}{p_{\mathrm{ini}}(x)}\right)^{(1-\alpha)}\ee^{(1-\alpha)D_{\alpha}(p_{\mathrm{ini}}||p_{\lambda_{0}}^{\mathrm{can}})} \nonumber \\
&=& P[\Gamma|x]p_{\alpha}(x), \label{thermoalpha}
\eeqa
where 
\beq
p_{\alpha}(x)=(p_{\mathrm{ini}}(x))^{\alpha}(p_{\lambda_{0}}^{\mathrm{can}}(x))^{1-\alpha}\ee^{(1-\alpha)D_{\alpha}(p_{\mathrm{ini}}||p_{\lambda_{0}}^{\mathrm{can}})} \label{localcanonicalalpha}
\eeq
is the canonical distribution with respect to the Hamiltonian $H_{\alpha}(p_{\mathrm{ini}})=(1-\alpha)H_{\lambda_{0}}+\alpha H(p_{\mathrm{ini}})$. Following a similar argument for the protocol $\alpha=0$, we can show that the explicit protocol is given by the following two steps: (1) Change the Hamiltonian of the system from $H_{\lambda_{0}}$ to $H_{\alpha}(p_{\mathrm{ini}})$ by keeping the initial distribution $p_{\mathrm{ini}}(x)$ fixed; (2) Change the Hamiltonian from $H_{\alpha}(p_{\mathrm{ini}})$ to $H_{\lambda_{1}}$ slowly in contact with the heat bath. Due to this interaction with the heat bath, the distribution $p_{\mathrm{ini}}$ thermalizes to the canonical distribution $p_{\alpha}$ at the beginning of the protocol (2). Note that the explicit protocol is given by mixing two protocols $\alpha=0$ and $\alpha=1$ by the ratio $\alpha:1-\alpha$. See Fig.~2 in the main text which describes a change in the energy levels in protocol (1). 

Note that the protocol (1) described above can be thought of as an ordinary adiabatic process if we can turn off the interaction between the system and the heat bath during the control of the Hamiltonian. If the heat bath is always in contact with the system, we need to quench the Hamiltonian instantaneously.

\section{\label{sec:noneqfinal} The case of a nonequilibrium final state}

In this section, we consider the case in which the final state is out of equilibrium. Recall that the lower bound of~(\ref{deee}) is satisfied if and only if the variance of $\alpha  \sigma-(1-\alpha) W$ is equal to zero, and the constant $\alpha$ takes a value between $[0,1]$. By setting the condition
\beq
a\sigma[\Gamma]-b\beta W[\Gamma]=b\beta\Delta F^{\mathrm{eq}}+c' , \label{dgg}
\eeq
we obtain a relation between forward and backward probability distributions:
\beq
\exp(c')=\frac{P[\Gamma]}{\tilde{P}[\Gamma^{\dagger}]}\left( \frac{p_{\mathrm{ini}}(x)}{p_{\lambda_{0}}^{\mathrm{can}}(x)}\right)^{-(1-\alpha)}\left( \frac{p_{\mathrm{fin}}(y)}{p_{\lambda_{1}}^{\mathrm{can}}(y)}\right)^{1-\alpha}, \label{dgbdcon}
\eeq
where $c'$ is a constant which will be determined later. We first identify an explicit condition on path probabilities which gives the lower bound of the trade-off relation~(4) in the main text by rewriting the above condition~(\ref{dgbdcon}):
\beq
P[\Gamma|x]p_{\alpha}(x)=\tilde{P}[\Gamma^{\dagger}|y]p_{\alpha}'(y) . \label{bdgbcona}
\eeq
Here, we use the definition of the canonical distribution $p_{\alpha}$ with respect to the Hamiltonian $H_{\alpha}$ given in Eq.~(\ref{localcanonicalalpha}). Similarly, we define the canonical distribution $p_{\alpha}'$ with respect to $H_{\alpha}(p_{\mathrm{fin}})$, where 
\beq
p_{\alpha}'(y)=(p_{\mathrm{fin}}(y))^{\alpha}(p_{\lambda_{1}}^{\mathrm{can}}(y))^{1-\alpha}\ee^{c'-(1-\alpha)D_{\alpha}(p_{\mathrm{ini}}||p_{\lambda_{0}}^{\mathrm{can}})} \label{finalalphacan}
\eeq
and $H_{\alpha}(p_{\mathrm{fin}})=(1-\alpha) H_{\lambda_{1}}+\alpha H(p_{\mathrm{fin}})$. Here, the Hamiltonian $H(p_{\mathrm{fin}})$ is defined such that the canonical distribution for $H(p_{\mathrm{fin}})$ with the inverse temperature $\beta$ is equal to $p_{\mathrm{fin}}$. Note that the constant $c'$ is determined by the normalization condition of the distribution~(\ref{finalalphacan}).

The condition~(\ref{bdgbcona}) is satisfied if the system Hamiltonian is slowly changed from $H_{\alpha}$ to $H_{\alpha}(p_{\mathrm{fin}})$. However, we should note that we cannot transform the distribution $p_{\alpha}'$ to $p_{\mathrm{fin}}$ in contact with a heat bath, because we do not have a time-reversal protocol of the thermalization process. Due to this asymmetry in time, we consider a modified protocol in which the support of the distribution $p_{\alpha}'(y)$ is restricted to $y\in Y$. By invoking the idea of thermo-majorization (Ref.~[18] in the main text) which is to be explained later, we determine $Y$ in which $p_{\alpha}'(y)$ can be transformed to $p_{\mathrm{fin}}(y)$ only by exchanging heat with the heat bath (i.e., vanishing work fluctuation). Then, the normalization condition fixes the constant $c'$:
\beq
c'=(1-\alpha)D_{\alpha}(p_{\mathrm{ini}}||p_{\lambda_{0}}^{\mathrm{can}})+(1-\alpha)d_{\alpha}(p_{\mathrm{fin}}||p_{\lambda_{1}}^{\mathrm{can}}) \label{dgconstant}
\eeq
and
\beq
(1-\alpha)d_{\alpha}(p_{\mathrm{fin}}||p_{\lambda_{1}}^{\mathrm{can}})=-\ln\sum_{y\in Y}(p_{\mathrm{fin}}(y))^{\alpha}(p_{\lambda_{1}}^{\mathrm{can}}(y))^{1-\alpha}.
\eeq
Then,
\beq
p_{\alpha}'=\frac{\ee^{-\beta H_{\alpha}^{Y}(p_{\mathrm{fin}})}}{Z_{\alpha}^{Y}(p_{\mathrm{fin}})}:=(p_{\mathrm{fin}})^{\alpha}(p_{\lambda_{1}}^{\mathrm{can}})^{1-\alpha}\ee^{(1-\alpha)d_{\alpha}(p_{\mathrm{fin}}||p_{\lambda_{1}}^{\mathrm{can}})},
\eeq
where $H_{\alpha}^{Y}(p_{\mathrm{fin}})$ is obtained by taking the Hamiltonian $H_{\alpha}(p_{\mathrm{fin}})$ and restricting its support to $Y$.

Now let us briefly review thermal operation and thermo-majorization which we use to determine $Y$. Let $\rho^{S}$ be a density matrix with no off-diagonal components in the energy eigenbasis of the system Hamiltonian $H^{S}$. Then, the thermal operation $\mathcal{E}$ is defined by the following map
\beq
\mathcal{E}(\rho^{S})=\Tr_{B}[U^{SB}(\rho^{S}\otimes\sigma^{B}_{\mathrm{can}})U^{\dagger SB}],
\eeq
where $\sigma^{B}_{\beta}$ is the canonical distribution of the heat bath with respect to the Hamiltonian $H^{B}$ with the inverse temperature $\beta$, and $\Tr_{B}$ denotes a partial trace over the degrees of freedom of the heat bath. We also require that the unitary operator $U^{SB}$ satisfies the energy conservation of the composite system:
\beq
[U^{SB},H^{S}+H^{B}]=0.
\eeq
Since the initial state does not have coherence, we can also think of $\mathcal{E}$ as a stochastic map with total energy conservation for a classical system. From the total energy conservation, the internal energy change $\Delta E$ of the system is equal to the heat $Q$ transfered from the heat bath to the system, i.e., the work is zero during this operation.

Next, let $\{E_{\alpha}^{p_{\mathrm{fin}}}(y)\}$ represent the eigenspectrum of the Hamiltonian $H_{\alpha}(p_{\mathrm{fin}})$. We arrange the final state $p_{\mathrm{fin}}(y)$ according to the following order:
\beq
\frac{p_{\mathrm{fin}}(y_{1})}{(Z_{\alpha}(p_{\mathrm{fin}}))^{-1}\ee^{-\beta E_{\alpha}^{p_{\mathrm{fin}}}(y_{1})}} \geq  \frac{p_{\mathrm{fin}}(y_{2})}{(Z_{\alpha}(p_{\mathrm{fin}}))^{-1}\ee^{-\beta E_{\alpha}^{p_{\mathrm{fin}}}(y_{2})}} \geq \cdots. \label{ordering}
\eeq
Then, we plot a convex curve (Lorenz curve~[31]) in the $(y,x)$ plane in which each point is given by
\beq
(p_{\mathrm{fin}},\frac{\ee^{-\beta H_{\alpha}(p_{\mathrm{fin}})}}{Z_{\alpha}(p_{\mathrm{fin}})})=((p_{\mathrm{fin}}(y_{1}),\frac{\ee^{-\beta E_{\alpha}^{p_{\mathrm{fin}}}(y_{1})}}{Z_{\alpha}(p_{\mathrm{fin}})}),(\sum_{i=1}^{2}p_{\mathrm{fin}}(y_{i}),\sum_{i=1}^{2}\frac{\ee^{-\beta E_{\alpha}^{p_{\mathrm{fin}}}(y_{i})}}{Z_{\alpha}(p_{\mathrm{fin}})}),(\sum_{i=1}^{3}p_{\mathrm{fin}}(y_{i}),\sum_{i=1}^{3}\frac{\ee^{-\beta E_{\alpha}^{p_{\mathrm{fin}}}(y_{i})}}{Z_{\alpha}(p_{\mathrm{fin}})}),\cdots, (1,1)), \label{curve}
\eeq
where $Z_{\alpha}(p_{\mathrm{fin}})=\sum_{i}\ee^{-\beta E_{\alpha}^{p_{\mathrm{fin}}}(y_{i})}$ is the partition function with respect to the Hamiltonian $H_{\alpha}(p_{\mathrm{fin}})$. Note that the ordering~(\ref{ordering}) ensures that the curve~(\ref{curve}) is convex. Then, it has been proven in Ref.~[18] in the main text that a state transformation from $p$ to $q$ is possible by thermal operation (with the system Hamiltonian $H_{\alpha}(p_{\mathrm{fin}})$) if and only if the curve $(q,\frac{\ee^{-\beta H_{\alpha}(p_{\mathrm{fin}})}}{Z_{\alpha}(p_{\mathrm{fin}})})$ is a subset of the curve $(p,\frac{\ee^{-\beta H_{\alpha}(p_{\mathrm{fin}})}}{Z_{\alpha}(p_{\mathrm{fin}})})$.

\begin{figure}[tbp]
\begin{center}
\includegraphics[width=.9\textwidth]{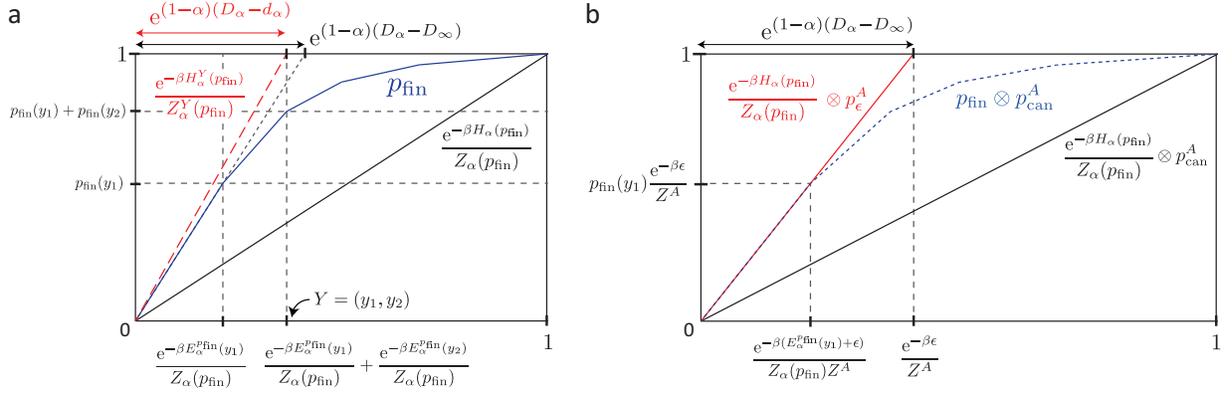}
\caption{(a) {\bf Lorenz curve and the definition of the support $Y$.} We plot $\{\sum_{i=1}^{k}\ee^{-\beta E_{\alpha}^{p_{\mathrm{fin}}}(y_{i})}/Z_{\alpha}(p_{\mathrm{fin}})\}_{k}$ and $\{\sum_{i=1}^{k}p(y_{i})\}_{k}$ in the $(x,y)$ plane, where the curve $p=\ee^{-\beta H_{\alpha}^{Y}(p_{\mathrm{fin}})}/Z_{\alpha}^{Y}(p_{\mathrm{fin}})$ is represented by the dashed line and $p=p_{\mathrm{fin}}$ is  represented by the solid curve. The dotted line is an extension of the line connecting two points: $(0,0)$ and $(\ee^{-\beta E_{\alpha}^{p_{\mathrm{fin}}}(y_{1})}/Z_{\alpha}(p_{\mathrm{fin}}),p_{\mathrm{fin}}(y_{1}))$. The support $Y$ is defined as the largest support such that the slope of the line $\ee^{-\beta H_{\alpha}^{Y}(p_{\mathrm{fin}})}/Z_{\alpha}^{Y}(p_{\mathrm{fin}})$ is equal to or larger than that of the dotted line. Then, the thermo-majorization criterion tells us that the local equilibrium state $\ee^{-\beta H_{\alpha}^{Y}(p_{\mathrm{fin}})}/Z_{\alpha}^{Y}(p_{\mathrm{fin}})$ can be transformed into $p_{\mathrm{fin}}$ via a thermal operation. (b) {\bf The case of introducing an auxiliary system.} By introducing an auxiliary system, the joint probability distribution $\ee^{-\beta H_{\alpha}(p_{\mathrm{fin}})}/Z_{\alpha}(p_{\mathrm{fin}})\otimes p^{A}_{\mathrm{\epsilon}}$ can be transformed into $p_{\mathrm{fin}}\otimes p^{A}_{\mathrm{can}}$ if the slope of the line $\ee^{-\beta H_{\alpha}(p_{\mathrm{fin}})}/Z_{\alpha}(p_{\mathrm{fin}})\otimes p^{A}_{\mathrm{\epsilon}}$ is equal to or larger than that of the line connecting two points: $(0,0)$ and $(\ee^{-\beta (E_{\alpha}^{p_{\mathrm{fin}}}(y_{1})+\epsilon)}/(Z_{\alpha}(p_{\mathrm{fin}})Z^{A}),p_{\mathrm{fin}}(y_{1})\ee^{-\beta\epsilon}/Z^{A})$. }
\label{fig:thermal}
\end{center}
\end{figure}

Now our task is to find a local canonical distribution $\ee^{-\beta H_{\alpha}^{Y}(p_{\mathrm{fin}})}/Z_{\alpha}^{Y}(p_{\mathrm{fin}})$ with support $Y$ such that by a thermal operation, it can be transformed to $p_{\mathrm{fin}}$. By plotting two curves $(p_{\mathrm{fin}},\frac{\ee^{-\beta H_{\alpha}(p_{\mathrm{fin}})}}{Z_{\alpha}(p_{\mathrm{fin}})})$ and $(\frac{\ee^{-\beta H_{\alpha}^{Y}(p_{\mathrm{fin}})}}{Z_{\alpha}^{Y}(p_{\mathrm{fin}})},\frac{\ee^{-\beta H_{\alpha}(p_{\mathrm{fin}})}}{Z_{\alpha}(p_{\mathrm{fin}})})$ as in Fig.~\ref{fig:thermal}. (a), we conclude that the slope of the latter curve should be larger than that of the former:
\beq
\frac{1}{\sum_{y\in Y}(Z_{\alpha}(p_{\mathrm{fin}}))^{-1}\ee^{-\beta E_{\alpha}^{p_{\mathrm{fin}}}(y)}} \geq \frac{p_{\mathrm{fin}}(y_{1})}{(Z_{\alpha}(p_{\mathrm{fin}}))^{-1}\ee^{-\beta E_{\alpha}^{p_{\mathrm{fin}}}(y_{1})}} . \label{Ycondition}
\eeq
We choose the largest support of $Y$ that satisfies the condition~(\ref{Ycondition}). We denote the eigenenergies $E_{p_{\mathrm{fin}}}(y)$ and $E_{\lambda_{1}}(y)$ corresponding to the Hamiltonians $H(p_{\mathrm{fin}})$ and $H_{\lambda_{1}}$. We also introduce the partition functions $Z(p_{\mathrm{fin}})=\sum_{y}\exp(-\beta E_{p_{\mathrm{fin}}}(y))$ and $Z_{\lambda_{1}}=\sum_{y}\exp(-\beta E_{\lambda_{1}}(y))$. It follows from $p_{\mathrm{fin}}(y)=\exp(-\beta E_{p_{\mathrm{fin}}}(y))/Z(p_{\mathrm{fin}})$ that the Renyi divergence between the final and canonical distributions is given by
\beq
(\alpha-1)D_{\alpha}(p_{\mathrm{fin}}||p_{\lambda_{1}}^{\mathrm{can}})=\ln\sum_{y}(p_{\mathrm{fin}}(y))^{\alpha}(p_{\lambda_{1}}^{\mathrm{can}}(y))^{1-\alpha}=\ln \frac{Z_{\alpha}(p_{\mathrm{fin}})}{(Z(p_{\mathrm{fin}}))^{\alpha}(Z_{\lambda_{1}})^{1-\alpha}}. \label{finalRenyia}
\eeq
Using Eq.~(\ref{finalRenyia}), the right-hand side of~(\ref{Ycondition}) can be expressed as
\beqa
\frac{Z_{\alpha}(p_{\mathrm{fin}})p_{\mathrm{fin}}(y_{1})}{\ee^{-\beta E_{\alpha}^{p_{\mathrm{fin}}}(y_{1})}}&=&\frac{Z_{\alpha}(p_{\mathrm{fin}})}{(Z(p_{\mathrm{fin}}))^{\alpha}(Z_{\lambda_{1}})^{1-\alpha}}\left(\frac{p_{\mathrm{fin}}(y_{1})}{p_{\lambda_{1}}^{\mathrm{can}}(y_{1})}\right)^{1-\alpha} \nonumber \\
&=&\exp\left[(1-\alpha)(D_{\infty}(p_{\mathrm{fin}}||p_{\lambda_{1}}^{\mathrm{can}})-D_{\alpha}(p_{\mathrm{fin}}||p_{\lambda_{1}}^{\mathrm{can}})) \right],
\eeqa
where
\beq
D_{\infty}(p||q)=\ln\max_{y}\frac{p(y)}{q(y)}
\eeq
is the Renyi divergence of order $\infty$. The left-hand side of~(\ref{Ycondition}) can be expressed as
\beq
\frac{Z_{\alpha}(p_{\mathrm{fin}})}{\sum_{y\in Y}\ee^{-\beta E_{\alpha}^{p_{\mathrm{fin}}}(y)}} =\exp\left[(1-\alpha)(d_{\alpha}(p_{\mathrm{fin}}||p_{\lambda_{1}}^{\mathrm{can}})-D_{\alpha}(p_{\mathrm{fin}}||p_{\lambda_{1}}^{\mathrm{can}})) \right].
\eeq
Now the condition~(\ref{Ycondition}) to determine $Y$ is rewritten in terms of the Reny divergences as
\beq
\exp\left[(1-\alpha)(d_{\alpha}(p_{\mathrm{fin}}||p_{\lambda_{1}}^{\mathrm{can}})-D_{\alpha}(p_{\mathrm{fin}}||p_{\lambda_{1}}^{\mathrm{can}}))\right] \geq \exp\left[(1-\alpha)(D_{\infty}(p_{\mathrm{fin}}||p_{\lambda_{1}}^{\mathrm{can}})-D_{\alpha}(p_{\mathrm{fin}}||p_{\lambda_{1}}^{\mathrm{can}})) \right]. \label{Yconddiv}
\eeq
We note that this condition can be expressed by using the Lorenz curve as shown in Fig.~\ref{fig:thermal}. (a). If the energy level $E_{\alpha}^{p_{\mathrm{fin}}}(y)$ is dense, we can choose $Y$ such that the equality condition of~(\ref{Yconddiv}) holds. Also, if we can attach an auxiliary system, we can tune the energy level of the total system in such a manner that the equality condition of~(\ref{Yconddiv}) holds as described in the next subsection.

Having established a method to determine $Y$, we can calculate the functional form of $\sigma$ and $W$ for the boundary by combining Eqs.~(\ref{dgbdcon}) and (\ref{dgconstant}):
\beqa
\sigma[\Gamma]&=&(1-\alpha)(\mathcal{D}_{\lambda_{0}}(x)-\mathcal{D}_{\alpha}-\mathcal{D}_{\lambda_{1}}(y)+d_{\alpha}), \label{sigmaalphaf} \\
\beta W[\Gamma]&=&-\beta \Delta F^{\mathrm{eq}}+(1-\alpha)(\mathcal{D}_{\alpha}-d_{\alpha})+\alpha(\mathcal{D}_{\lambda_{0}}(x)-\mathcal{D}_{\lambda_{1}}(y)),
\eeqa
where $\mathcal{D}_{\lambda_{1}}(y)=\ln p_{\mathrm{fin}}(y)-\ln p_{\lambda_{1}}^{\mathrm{can}}(y)$ is the distance between the final state and the canonical distribution.

Let us now derive the trade-off relation between work fluctuation and dissipation for arbitrary initial and final states. We fix the amount of work fluctuation as
\beq
\sqrt{\av{W^2}-\av{W}^{2}}=\alpha\sqrt{\av{(\Delta \mathcal{F})^{2}}-\av{\Delta \mathcal{F}}^{2} }, \label{workflucfixeddg}
\eeq
where $\alpha\in[0,1]$ is a constant. We use exactly the same method that we use in deriving~(\ref{sup:sigmaflucdistradeoff}) except that $\sigma_{\alpha}[\Gamma]$ in the present case is given by Eq.~(\ref{sigmaalphaf}). Then, for the fixed work fluctuation~(\ref{workflucfixeddg}), the bounds on dissipation and the average work take the following forms:
\beqa
\av{\sigma} &\geq& (1-\alpha)(D(p_{\mathrm{ini}}||p_{\lambda_{0}}^{\mathrm{can}})-D_{\alpha}(p_{\mathrm{ini}}||p_{\lambda_{0}}^{\mathrm{can}})) +(1-\alpha)(d_{\alpha}(p_{\mathrm{fin}}||p_{\lambda_{1}}^{\mathrm{can}})-D(p_{\mathrm{fin}}||p_{\lambda_{1}}^{\mathrm{can}})) , \label{disbounddg} \\
\av{W}&\leq& -\alpha \av{\Delta \mathcal{F}} -(1-\alpha)(f_{\alpha}(p_{\mathrm{fin}})(p_{\mathrm{fin}})-F_{\alpha}(p_{\mathrm{fin}}). \label{workbounddg}
\eeqa

\subsection{The case of introducing an auxiliary system}
Here, we consider a different setup of a two-level auxiliary system $A$, whose eigenenergies are given by $E^{A}(0)=0,\ E^{A}(1)=\epsilon$. We construct a protocol which gives the equality condition of~(\ref{Yconddiv}) in this setup. For simplicity, we first consider the case of $\alpha=0$. We consider a transition from $p^{\mathrm{can}}_{\lambda_{1}} \otimes p^{A}_{\epsilon}$ to $p_{\mathrm{fin}}\otimes p^{A}_{\mathrm{can}}$ via a thermal operation, where $p^{A}_{\epsilon}(0)=0, p^{A}_{\epsilon}(1)=1$. Note that the support of the joint distribution $p^{\mathrm{can}}_{\lambda_{1}} \otimes p^{A}_{\epsilon}$ is restricted to ${(y,1)}$. Then, 
\beq
d_{0}(p_{\mathrm{fin}}\otimes p^{A}_{\mathrm{can}}||p^{\mathrm{can}}_{\lambda_{1}} \otimes p^{A}_{\epsilon})=\ln p^{A}_{\mathrm{can}}(1)=\beta(\epsilon -F^{A}).
\eeq
If we choose $\epsilon$ such that 
\beq
D_{\infty}(p_{\mathrm{fin}}||p_{\lambda_{1}}^{\mathrm{can}})=\beta(\epsilon -F^{A})=d_{0}(p_{\mathrm{fin}}\otimes p^{A}_{\mathrm{can}}||p^{\mathrm{can}}_{\lambda_{1}} \otimes p^{A}_{\epsilon}),
\eeq
the lower bound of the trade-off relation~(\ref{disbounddg}) in the case of $\alpha=0$ is given by
\beq
\av{\sigma} \geq D(p_{\mathrm{ini}}||p_{\lambda_{0}}^{\mathrm{can}})-D_{0}(p_{\mathrm{ini}}||p_{\lambda_{0}}^{\mathrm{can}}) +D_{\infty}(p_{\mathrm{fin}}||p_{\lambda_{1}}^{\mathrm{can}})-D(p_{\mathrm{fin}}||p_{\lambda_{1}}^{\mathrm{can}}).
\eeq
Here, we note that the quantity $\epsilon -F^{A}$ is equal to the work cost of creating $p^{A}_{\epsilon}$ starting from $p^{A}_{\mathrm{can}}$.

Next, let us consider the case of general $\alpha$. In this case, we consider a transition from $\exp(-\beta H_{\alpha}(p_{\mathrm{fin}}))/Z_{\alpha}(p_{\mathrm{fin}}) \otimes p^{A}_{\epsilon}$ to $p_{\mathrm{fin}}\otimes p^{A}_{\mathrm{can}}$ via a thermal operation. This operation is possible if $\epsilon$ satisfies
\beq
\exp\left[(1-\alpha)(D_{\infty}(p_{\mathrm{fin}}||p_{\lambda_{1}}^{\mathrm{can}})-D_{\alpha}(p_{\mathrm{fin}}||p_{\lambda_{1}}^{\mathrm{can}})) \right]=\frac{p_{\mathrm{fin}}(y_{1})}{\ee^{-\beta E_{\alpha}^{p_{\mathrm{fin}}}(y_{1})}/Z_{\alpha}(p_{\mathrm{fin}})}= \frac{1}{\ee^{-\beta \epsilon}/Z^{A}}, \label{qubitslope}
\eeq
where $Z^{A}=1+\ee^{-\beta \epsilon}$ is the partition function of $A$ (see Fig.~\ref{fig:thermal}. (b)).
By noting that the initial and final probability distributions of the ancillary system are given by  $p^{A}_{\epsilon}$ and $p^{A}_{\mathrm{can}}$, Eq.~(\ref{dgbdcon}) takes the form
\beq
\exp(c')=\frac{P[\Gamma] p^{A}_{\epsilon}(z)}{\tilde{P}[\Gamma^{\dagger}]p^{A}_{\mathrm{can}}(z) }\left( \frac{p_{\mathrm{ini}}(x)}{p_{\lambda_{0}}^{\mathrm{can}}(x)}\right)^{-(1-\alpha)}\left( \frac{p_{\mathrm{fin}}(y)}{p_{\lambda_{1}}^{\mathrm{can}}(y)}\right)^{1-\alpha}. 
\eeq
Using the following expression
\beq
\frac{\exp(-\beta E_{\alpha}^{p_{\mathrm{fin}}}(y))}{Z_{\alpha}(p_{\mathrm{fin}})}=(p_{\mathrm{fin}}(y))^{\alpha}(p_{\lambda_{1}}^{\mathrm{can}}(y))^{1-\alpha}\ee^{(1-\alpha)D_{\alpha}(p_{\mathrm{fin}}||p_{\lambda_{1}}^{\mathrm{can}})},
\eeq
the constant $c'$ is determined by
\beq
c'=(1-\alpha)D_{\alpha}(p_{\mathrm{ini}}||p_{\lambda_{0}}^{\mathrm{can}})+(1-\alpha)D_{\alpha}(p_{\mathrm{fin}}||p_{\lambda_{1}}^{\mathrm{can}})-\ln p^{A}_{\mathrm{can}}(1). \label{cqubit}
\eeq
Combining Eqs.~(\ref{qubitslope}) and (\ref{cqubit}), we obtain
\beq
c'=(1-\alpha)D_{\alpha}(p_{\mathrm{ini}}||p_{\lambda_{0}}^{\mathrm{can}})+(1-\alpha)D_{\infty}(p_{\mathrm{fin}}||p_{\lambda_{1}}^{\mathrm{can}}).
\eeq
If the energy level of the auxiliary system satisfies Eq.~(\ref{qubitslope}), the lower bound of the trade-off relation~(\ref{disbounddg}) is lowered and takes the form
\beq
\av{\sigma} \geq (1-\alpha)(D(p_{\mathrm{ini}}||p_{\lambda_{0}}^{\mathrm{can}})-D_{\alpha}(p_{\mathrm{ini}}||p_{\lambda_{0}}^{\mathrm{can}})) +(1-\alpha)(D_{\infty}(p_{\mathrm{fin}}||p_{\lambda_{1}}^{\mathrm{can}})-D(p_{\mathrm{fin}}||p_{\lambda_{1}}^{\mathrm{can}})) .
\eeq

\subsection{Extractable work for each step of the protocol that achieve the lower bound of the trade-off relation}
Here, we calculate the work and dissipation for each step of the explicit protocol. First, let us denote the canonical distributions as
\beqa
\frac{\ee^{-\beta H_{\alpha}(p_{\mathrm{ini}})}}{Z_{\alpha}(p_{\mathrm{ini}})}&=&(p_{\mathrm{ini}})^{\alpha}(p^{\mathrm{can}}_{\lambda_{0}})^{1-\alpha}\ee^{(1-\alpha)D_{\alpha}}  ,\\
\frac{\ee^{-\beta H_{\alpha}^{Y}(p_{\mathrm{fin}})}}{Z_{\alpha}^{Y}(p_{\mathrm{fin}})}&=&(p_{\mathrm{fin}})^{\alpha}(p^{\mathrm{can}}_{\lambda_{1}})^{1-\alpha}\ee^{(1-\alpha)d_{\alpha}} ,
\eeqa
where
\beq
H_{\alpha}^{Y}(p_{\mathrm{fin}})=\sum_{y\in Y}E_{\alpha}^{p_{\mathrm{fin}}}(y)
\eeq
is the local Hamiltonian whose support is restricted to $Y$ and $Z_{\alpha}^{Y}(p_{\mathrm{fin}})=\sum_{y\in Y}\exp(-\beta E_{\alpha}^{p_{\mathrm{fin}}}(y))$ is the corresponding partition function. Let us denote $(p,H)$ as the set of the distribution $p$ and the Hamiltonian $H$. We also denote the nonequilibrium free energy as 
\beq
\mathcal{F}(p,H)=F^{\mathrm{eq}}+\beta^{-1}D(p||\frac{\ee^{-\beta H}}{Z}),
\eeq
where $F^{\mathrm{eq}}$ is the equilibrium free energy with respect to the Hamiltonian $H$. The explicit protocol that achieves the lower bound of (2), (4) and (5) is given by
\begin{enumerate}
\item[1.] Quench process $(p_{\mathrm{ini}},H_{\lambda_{0}})\rightarrow(p_{\mathrm{ini}},H_{\alpha}(p_{\mathrm{ini}}))$. 

The extractable work during this quench process is given by 
\beqa
\av{W}_{1}&=&\sum_{x}p_{\mathrm{ini}}(x)(E_{\lambda_{0}}(x)-E_{\alpha}^{p_{\text{ini}}}(x)) \nonumber \\
&=&\alpha\sum_{x}p_{\mathrm{ini}}(x)(E_{\lambda_{0}}(x)-E_{p_{\mathrm{ini}}}(x)) \nonumber \\
&=&\alpha\av{\mathcal{F}_{\lambda_{0}}}-\alpha F^{\text{eq}}(p_{\mathrm{ini}}).
\eeqa

\item[2.] Thermalization process $(p_{\mathrm{ini}},H_{\alpha}(p_{\mathrm{ini}}))\rightarrow(\frac{\ee^{-\beta H_{\alpha}(p_{\mathrm{ini}})}}{Z_{\alpha}(p_{\mathrm{ini}})},H_{\alpha}(p_{\mathrm{ini}}))$. 

The extractable work vanishes and the dissipation is given by the nonequilibrium free-energy difference:
\beqa
\beta^{-1}\av{\sigma}_{2}&=&\mathcal{F}(p_{\mathrm{ini}},H_{\alpha}(p_{\mathrm{ini}}))-\mathcal{F}(\ee^{-\beta H_{\alpha}(p_{\mathrm{ini}})}/Z_{\alpha}(p_{\mathrm{ini}}),H_{\alpha}(p_{\mathrm{ini}}))=(1-\alpha)(\av{\mathcal{F}_{\lambda_{0}}}-\mathcal{F}_{\alpha}) \nonumber \\
&=&\beta^{-1}D(p_{\mathrm{ini}}||\ee^{-\beta H_{\alpha}(p_{\mathrm{ini}})}/Z_{\alpha}(p_{\mathrm{ini}}))=(1-\alpha)\Delta D_{\alpha}(p_{\mathrm{ini}}||p^{\mathrm{can}}_{\lambda_{0}}).
\eeqa

\item[3.a.] Quasi-static process $(\frac{\ee^{-\beta H_{\alpha}(p_{\mathrm{ini}})}}{Z_{\alpha}(p_{\mathrm{ini}})},H_{\alpha}(p_{\mathrm{ini}}))\rightarrow(\frac{\ee^{-\beta H_{\alpha}^{Y}(p_{\mathrm{fin}})}}{Z_{\alpha}^{Y}(p_{\mathrm{fin}})},H_{\alpha}^{Y}(p_{\mathrm{fin}}))$.

The extractable work is equal to the equilibrium free-energy difference
\beqa
\av{W}_{3}&=&F^{\mathrm{eq}}\left(\frac{\ee^{-\beta H_{\alpha}(p_{\mathrm{ini}})}}{Z_{\alpha}(p_{\mathrm{ini}})}\right)-F^{\mathrm{eq}}\left(\frac{\ee^{-\beta H_{\alpha}^{Y}(p_{\mathrm{fin}})}}{Z_{\alpha}^{Y}(p_{\mathrm{fin}})}\right)\nonumber \\
&=&(1-\alpha)\mathcal{F}_{\alpha}(p_{\mathrm{ini}})+\alpha F^{\text{eq}}(p_{\mathrm{ini}})-(1-\alpha)f_{\alpha}(p_{\mathrm{fin}}) - \alpha F^{\text{eq}}(p_{\mathrm{fin}}).
\eeqa

\item[3.b.] Quench process $(\frac{\ee^{-\beta H_{\alpha}^{Y}(p_{\mathrm{fin}})}}{Z_{\alpha}^{Y}(p_{\mathrm{fin}})},H_{\alpha}^{Y}(p_{\mathrm{fin}}))\rightarrow (\frac{\ee^{-\beta H_{\alpha}^{Y}(p_{\mathrm{fin}})}}{Z_{\alpha}^{Y}(p_{\mathrm{fin}})},H_{\alpha}(p_{\mathrm{fin}}))$.

The extractable work and the dissipation both vanish during this process.

\item[4.] Thermal operation $(\frac{\ee^{-\beta H_{\alpha}^{Y}(p_{\mathrm{fin}})}}{Z_{\alpha}^{Y}(p_{\mathrm{fin}})},H_{\alpha}(p_{\mathrm{fin}}))\rightarrow (p_{\mathrm{fin}},H_{\alpha}(p_{\mathrm{fin}}))$.

The extractable work vanishes and the dissipation is given by the nonequilibrium free-energy difference:
\beqa
\beta^{-1}\av{\sigma}_{4}&=&\mathcal{F}\left(\frac{\ee^{-\beta H_{\alpha}^{Y}(p_{\mathrm{fin}})}}{Z_{\alpha}^{Y}(p_{\mathrm{fin}})},H_{\alpha}(p_{\mathrm{fin}})\right)-\mathcal{F}(p_{\mathrm{fin}},H_{\alpha}(p_{\mathrm{fin}}))  \nonumber \\
&=&(1-\alpha)(f_{\alpha}(p_{\mathrm{fin}})-\av{\mathcal{F}_{\lambda_{1}}})=(1-\alpha)\beta^{-1}\Delta d_{\alpha}(p_{\mathrm{fin}}||p^{\mathrm{can}}_{\lambda_{1}}).
\eeqa

\item[5.] Quench process $(p_{\mathrm{fin}},H_{\alpha}(p_{\mathrm{fin}}))\rightarrow (p_{\mathrm{fin}},H_{\lambda_{1}})$. 

The extractable work is given by
\beq
\av{W}_{5}=\sum_{y}p_{\mathrm{fin}}(y)(E_{\alpha}^{p_{\mathrm{fin}}}(y)-E_{\lambda_{1}}(y))=-\alpha\av{\mathcal{F}_{\lambda_{1}}}+\alpha F^{\mathrm{eq}}(p_{\mathrm{fin}}).
\eeq

\end{enumerate}

By combining the extractable work and the dissipation given above, the lower bounds of~(4) and (5) are obtained:
\beqa
\av{\sigma}_{2}+\av{\sigma}_{4}&=&(1-\alpha)\Delta D_{\alpha}(p_{\mathrm{ini}}||p^{\mathrm{can}}_{\lambda_{0}})+(1-\alpha)\Delta d_{\alpha}(p_{\mathrm{fin}}||p^{\mathrm{can}}_{\lambda_{1}}), \\
\av{W}_{1}+\av{W}_{3}+\av{W}_{5}&=&\alpha(\av{\mathcal{F}_{\lambda_{0}}}-\av{\mathcal{F}_{\lambda_{1}}})-(1-\alpha)(f_{\alpha}(p_{\mathrm{fin}})-\mathcal{F}_{\alpha}(p_{\mathrm{ini}})).
\eeqa

By noting that $f_{\alpha}(q) \geq \mathcal{F}_{\infty}(q) \geq  \mathcal{F}_{\alpha}(q)$, extractable work starting from a nonequilibrium state is always smaller than the work cost of preparing a nonequilibrium final state starting from equilibrium, as shown in the case of $\alpha=0$ in Ref.~[18]. This relation holds for a general $\alpha$ with fixed work fluctuation except for the reversible regime $\alpha=1$ because of the asymmetry of the protocol.




\section{Detailed description of the numerical simulations}
\subsection{Numerical simulation in Fig. 1. (b)}
A numerical simulation is done in a five-level system, and we plotted the dissipation versus the work fluctuation in Fig. 1. (b) in the main text. We choose the initial distribution as $p_{\mathrm{ini}}(x)=\{0.7,0.2,0.075,0.025,0\}$. We also set the canonical distribution with respect to the initial Hamiltonian $H_{\lambda_{0}}$ as $p^{\mathrm{can}}_{\lambda_{0}}(x)=\{0.5,0.3,0.1,0.075,0.025\}$. The numerical simulation is carried out by randomly generating a quenched Hamiltonian $H_{\mathrm{quench}}$. During this quenching process, work is put in or extracted from the system, given by $W(x)=E_{\lambda_{0}}(x)-E_{\mathrm{quench}}(x)$. Using this expression of work, we can calculate the work fluctuation along this process. After the quench of the Hamiltonian, we consider an ideal thermalization process in which the thermalized state is given by the canonical distribution $p^{\mathrm{can}}_{\mathrm{quench}}(x)$ with respect to the quenched Hamiltonian $H_{\mathrm{quench}}$. The energy dissipated during this process is quantified by $\sigma(x)=\ln p_{\mathrm{ini}}(x)-\ln p_{\mathrm{quench}}^{\mathrm{can}}(x)$. After the thermalization process, we consider an isothermal expansion by changing the Hamiltonian from $H_{\mathrm{quench}}$ to $H_{\lambda_{1}}$; then, the work fluctuation and the dissipation vanish during this process. Note that this idealized thermalization and isothermal processes are enough to explore the lower bound of the trade-off relation for the thermalized final state. We plot the following quantities for each quenched Hamiltonian in Fig. 1. (b):
\beqa
\av{W^{2}}-\av{W}^{2}&=& \sum_{x}p_{\mathrm{ini}}(x)(W(x))^{2}-\left(\sum_{x}p_{\mathrm{ini}}(x)W(x)\right)^{2}, \\
\av{\sigma}&=&\sum_{x}p_{\mathrm{ini}}(x)\sigma(x).
\eeqa

In Fig. 1. (b), we choose $p_{\mathrm{fin}}(y)=\{0.6,0.2,0.1,0.075,0.025\}$ and $p^{\mathrm{can}}_{\lambda_{1}}(y)=\{0.5,0.2,0.15,0.1,0.05\}$ to calculate the lower bound of the trade-off relation for a target final state $p_{\mathrm{fin}}$. Note that the support $Y$ in the definition of $d_{\alpha}(p_{\mathrm{fin}}||p^{\mathrm{can}}_{\lambda_{1}})$ changes from $Y=\{0,1\}$ to $Y=\{0,1,2\}$ around $\alpha=0.13$ and from $Y=\{0,1,2\}$ to $Y=\{0,1,2,3\}$ around $\alpha=0.81$.

\subsection{Numerical simulation in Fig. 1 (c)}
We consider a Szilard engine-like information heat engine in a single electron box in Fig. 1 (c) in the main text, following Ref.~[15] in the main text (we also follow the parameters of numerical simulation described in the supplementary material of Ref.~[15] in the main text). We consider a two-leveled system whose initial distribution is given by $p^{S}(n)=\{\half,\half\}$, where $n=\{0,1\}$ labels the state of the system. The internal energy of the system is given by $E(n)=E_{\mathrm{c}}(n-n_{\mathrm{g}})^{2}$, where $E_{\mathrm{c}}$ is the total charging energy, $n_{\mathrm{g}}$ is a control parameter which can tune the energy level of the system by changing the gate voltage. The initial and final Hamiltonians are given by setting $n_{\mathrm{g}}=\half$.

We consider a measurement of the system, where the joint probability distribution of the state of the system being $n$ and the measurement outcome being $m$ is given by $P(n,m)=(1-\epsilon)/2$ for $m=n$ and $P(n,m)=\epsilon/2$ for $m\neq n$. Here, we set the error probability as $\epsilon=0.02$. For example, the postmeasurement state conditioned on the measurement outcome $m=0$ is given by $P(n|m=0)=\{1-\epsilon,\epsilon\}$. We change the control parameter $n_{\mathrm{g}}$ depending on the measurement outcome. One typical example (for $m=0$) is to change $n_{\mathrm{g}}$ to $n_{\text{quench}}=0.349$ instantaneously, followed by a slow return to the degeneracy point:
\beq
n_{\mathrm{g}}(t)=n_{\text{quench}}+(0.5-n_{\text{quench}})\frac{\log(1+t)}{\log(1+T)},
\eeq
where $t$ is the time and $T$ is the total time needed to complete the feedback control. Note that $n_{\text{quench}}$ is determined from the condition that the canonical distribution with respect to the Hamiltonian with $n_{\mathrm{g}}=n_{\text{quench}}$ is equal to the distribution of the postmeasurement state.

The probability distribution during the feedback control is numerically calculated by using the following master equation
\beqa
\del{P(n=0,t)}{t}&=&-\Gamma_{0\rightarrow 1}(t)P(n=0,t)+\Gamma_{1\rightarrow 0}(t)P(n=1,t), \nonumber \\
\del{P(n=1,t)}{t}&=&-\Gamma_{1\rightarrow 0}(t)P(n=1,t)+\Gamma_{0\rightarrow 1}(t)P(n=0,t),
\eeqa
where $P(n,t)$ is the probability distribution of the system being $n$ at time $t$, and $\Gamma_{0\rightarrow 1}(t)$ is the tunneling rate of the single-electron box at time $t$ for the transition $0\rightarrow 1$ using the expression given in Ref.~[15] in the main text. The work is determined by the energy change of the system by changing the external parameter $n_{\mathrm{g}}(t)$:
\beq
W(n,t)=E_{t}(n)-E_{t+\Delta t}(n),
\eeq
where $E_{t}(n)=E_{\mathrm{c}}(n-n_{\mathrm{g}}(t))^{2}$ is the energy of the system at time $t$. The average extractable work from the system and work fluctuation are given by
\beqa
\av{W}&=&\sum_{i,n} W(n, i\Delta t)P(n,i\Delta t), \\
\av{W^{2}}-\av{W}^{2}&=& \sum_{i}\left\{ \sum_{n} W(n, i\Delta t)^{2}P(n,i\Delta t) - \left(\sum_{n}W(n, i\Delta t)P(n,i\Delta t)\right)^{2}    \right\},
\eeqa
where $\Delta t$ is the discretized time step. Different types of feedback protocols are obtained by changing $n_{\mathrm{quench}}$, $T$ and the functional form of $n_{\mathrm{g}}(t)$. For each feedback protocol, we calculate the dissipation and the work fluctuation, and plot them in Fig. 1 (c).

\end{document}